\documentclass[preprint,showpacs,aps,prl,superscriptaddress,letterpaper]{revtex4}
\usepackage{graphicx}
\usepackage{dcolumn}
\usepackage{amsmath}
\usepackage{epsfig}
\long\def\inst#1{\par\nobreak\kern 4pt\nobreak
    {\itshape #1}\par\vskip 10pt plus 3pt minus 3pt}
\RequirePackage{xspace}

\def\babar{\mbox{\slshape B\kern-0.1em{\smaller A}\kern-0.1em
    B\kern-0.1em{\smaller A\kern-0.2em R}}}
\def\Kbar    {\kern 0.18em\overline{\kern -0.18em K}{}\xspace}

\def\Kz      {\ensuremath{K^0}\xspace}
\def\Kzb     {\ensuremath{\Kbar^0}\xspace}
\def\KzKzb   {\ensuremath{\Kz {\kern -0.16em \Kzb}}\xspace}

\def\Ks     {\ensuremath{K_S}\xspace}
\def\Kl     {\ensuremath{K_L}\xspace}
\def\KsKs   {\ensuremath{\Ks {\kern -0.16em \Ks}}\xspace}
\def\KlKl   {\ensuremath{\Kl {\kern -0.16em \Kl}}\xspace}
\def\KsKl   {\ensuremath{\Ks {\kern -0.16em \Kl}}\xspace}
\def\KlKs   {\ensuremath{\Kl {\kern -0.16em \Ks}}\xspace}
\def\Dbar    {\kern 0.18em\overline{\kern -0.18em D}{}\xspace}
\def\Dz      {\ensuremath{D^0}\xspace}
\def\Dzb     {\ensuremath{\Dbar^0}\xspace}
\def\DzDzb   {\ensuremath{\Dz {\kern -0.16em \Dzb}}\xspace}
\newcommand{\DD}{\ensuremath{D\Dbar}\xspace}
\newcommand{\DsP}{\ensuremath{D_s^+}\xspace}
\newcommand{\DsM}{\ensuremath{D_s^-}\xspace}
\newcommand{\DspDsm}{\ensuremath{\DsP {\kern -0.16em \DsM}}\xspace}
\newcommand{\Dp}{\ensuremath{D^+}\xspace}
\newcommand{\Dm}{\ensuremath{D^-}\xspace}
\newcommand{\DpDm}{\ensuremath{\Dp {\kern -0.16em \Dm}}\xspace}

\newcommand{\phett}{$J/\psi\to\phi\eta(\eta^\prime)$}
\def\Bbar    {\kern 0.18em\overline{\kern -0.18em B}{}\xspace}

\def\Bz      {\ensuremath{B^0}\xspace}
\def\Bzb     {\ensuremath{\Bbar^0}\xspace}
\def\BzBzb   {\ensuremath{\Bz {\kern -0.16em \Bzb}}\xspace}
\def\Bu      {\ensuremath{B^+}\xspace}
\def\Bub     {\ensuremath{B^-}\xspace}

\def\BpBm    {\ensuremath{\Bu {\kern -0.16em \Bub}}\xspace}

\def\Dp      {\ensuremath{D^+}\xspace}

\newcommand{\optbar}[1]{\shortstack{{\tiny (\rule[.4ex]{1em}{.1mm})}
  \\ [-.7ex] $#1$}}
\def\BorBbar    {\kern 0.18em\optbar{\kern -0.18em B}{}\xspace}
\def\DorDbar    {\kern 0.18em\optbar{\kern -0.18em D}{}\xspace}
\def\KorKbar    {\kern 0.18em\optbar{\kern -0.18em K}{}\xspace}

\def\pep2{PEP-II}
\mathchardef\Upsilon="7107
\def\Y#1S{\ensuremath{\Upsilon{(#1S)}}\xspace}

\begin{document}

\title{\large \bfseries \boldmath $\eta$ and $\eta^\prime$ Physics at BES-III
}
\author{Hai-Bo Li}\email{lihb@ihep.ac.cn}
\affiliation{Institute of High Energy Physics, P.O.Box 918,
Beijing  100049, China}


\date{\today}


\begin{abstract}

Decays of both $\eta$ and $\eta^\prime$ provide very useful
information in our understanding of low-energy QCD, and
experimental signatures for these decays would be extremely
helpful at BES-III. The rare decays of the $\eta$ and
$\eta^\prime$ mesons could serve as a low-energy test of the
Standard Model and its beyond.  The sensitivities of the
measurements of $\eta$ and $\eta^\prime$ decays are discussed at
BES-III, in which the $\eta$ and $\eta^\prime$ mesons are produced
in the $\psi$ decays.

\end{abstract}


\maketitle

\section{Introduction}
  Quantum chromodynamics (QCD), which is a field theory of strong
  interaction, cannot be directly applied in the low energy since
  the strong coupling $\alpha_s$ is large~\cite{gross,politzer}. In
  this case one must resort to alternative model-independent
  approaches, such as lattice QCD, chiral perturbation theory
  (ChPT). The $\eta$ and $\eta^\prime$ mesons play an important role in
  understanding the low energy QCD. They are isoscalar members of
  the nonet of the lightest pseudoscalar mesons.
  Decays of $\eta$ and $\eta^\prime$ are investigated within
  a $U(3)$ chiral unitary approach based on the ChPT~\cite{chpt1,chpt2,chpt3}.
  Precision measurements on $\eta$ and $\eta^\prime$ would be very
  helpful, and provide useful information in our understanding of
  low energy QCD. Especially, the $\eta \rightarrow 3 \pi$ process
  is very interesting to verify the description of isospin
  violation in ChPT. The situation has never been clear on this
  mechanism. A precise analysis on the corresponding Dalitz plot
  parameters would be very helpful.

  The rare decays of $\eta$ and $\eta^\prime$ mesons also serve as
  a test of the Standard Model (SM) at low energy. In this paper,
  we are trying to give a review of $\eta$
  and $\eta^\prime$ physics at BES-III by using the decay of
  $\psi$.

\section{The production of $\eta$ and $\eta^\prime$ at BES-III}
Beginning in mid-2008, the BEPC-II/BES-III was operated at
center-of-mass (CM) energies corresponding to $\sqrt{s} = 2.0 -4.6
$ GeV. The design luminosity over this energy region  ranges from
$1\times 10^{33}$cm$^{-2}$s$^{-1}$ down to about $0.6 \times
10^{33}$cm$^{-2}$s$^{-1}$~\cite{bepcii}, yielding around 5
fb$^{-1}$ each year at $\psi(3770)$ above $\DspDsm$ threshold and
3 fb$^{-1}$ at $J/\psi$ peak in one year's running with full
luminosity~\cite{bepcii}. These integrated luminosities
correspond to samples of 2.0 million $\DspDsm$, 30 million $\DD$
pairs and $10\times 10^9$ $J/\psi$ decays. Table~\ref{tab:lum}
summarizes the data set per year at BES-III~\cite{haibo2006}. In
this paper, the sensitivity studies are based on 3 fb$^{-1}$
luminosity at the $J/\psi$ or $\psi(2S)$ peak for $\eta$ and
$\eta^\prime$ physics.
\begin{table}[htbp]
\centering
  \caption{$\tau$-Charm productions at BEPC-II in one year's running
($10^7s$).} {\begin{tabular}{@{}lll} \toprule  Data Sample&
Central-of-Mass  & \#Events \\
& (MeV) & per year  \\
\hline
$J/\psi$ &  3097     & $10\times 10^9$\\
$\tau^+\tau^-$   & 3670  & $12\times 10^6$ \\
$\psi(2S)$ & 3686  & $3.0\times 10^9$ \\
$\DzDzb$ & 3770  & $18\times 10^6$ \\
$\DpDm$ & 3770  & $14\times 10^6$ \\
$\DspDsm$ & 4030  & $1.0\times 10^6$ \\
$\DspDsm$ & 4170  & $2.0\times 10^6$ \\
\hline
\end{tabular}\label{tab:lum}}
\end{table}

In tables~\ref{tab:eta} and~\ref{tab:etap}, the dominant decay
channels of $J/\psi$ or $\psi(2S)$ decaying into final states
involving $\eta$ or $\eta^\prime$ mesons are listed. With one
year's luminosity at $J/\psi$ ($\psi(2S)$) peak, we expect to
obtain about 63 million $\eta$ decays and 61 million $\eta^\prime$
decays, respectively. At BEPC-II,  the background is small and the
event topology is simple comparing to the other experiments. The
decays of $\eta$ and $\eta^\prime$ can be studied with these clean
samples, especially, it will be one of the best place to study
$\eta^\prime$ physics.

\begin{table}[htbp]
\centering
  \caption{The production of $\eta$ meson at BES-III in charmonium decays by
assuming $10\times 10^9$ J$/\psi$ and $3.0\times 10^9$ $\psi(2S)$
events can be collected per year.} {\begin{tabular}{@{}lll}
\toprule Decay mode &
Combined branching fraction  & \#Events \\
& ($\times 10^{-4}$) &  per year  \\
\hline
$J/\psi \rightarrow \gamma \eta $ &  $(9.8\pm 1.0)$     & 9.8 $\times 10^6$ \\
$J/\psi \rightarrow \phi \eta (\phi \rightarrow K^+K^-)$ &  $(3.69\pm 0.39)$  & 3.69 $\times 10^6$ \\
$J/\psi \rightarrow \omega \eta (\omega \rightarrow \pi^+\pi^-\pi^0) $ &  $(15.5 \pm 1.8)$  & 15.5$\times 10^6$  \\
$J/\psi \rightarrow \rho \eta (\rho \rightarrow \pi^+\pi^-) $ &  $(1.93 \pm 0.23)$  & $1.93\times 10^6$ \\
$J/\psi \rightarrow p\bar{p} \eta  $ &  $(20.9\pm 1.8)$ & 20.9$\times 10^6$ \\
$\psi(2S) \rightarrow \eta J/\psi (J/\psi \rightarrow l^+l^-) $ & $(37.5\pm 0.8)$  & 11.25$\times 10^6$\\
\hline
 Total &   & 63.1$\times 10^6$ \\\hline
\end{tabular}\label{tab:eta}}
\end{table}

\begin{table}[htbp]
\centering
  \caption{The production of $\eta^\prime$ meson at BES-III in charmonium
decays by assuming $10\times 10^9$ J$/\psi$ and $3.0\times 10^9$
$\psi(2S)$ events can be collected per year.}
{\begin{tabular}{@{}lll} \toprule Decay mode &
Combined branching fraction  & \#Events \\
& ($\times 10^{-4}$) &  per year  \\
\hline
$J/\psi \rightarrow \gamma \eta^\prime $ &  $(47.1\pm 2.7)$     & 47.1 $\times 10^6$ \\
$J/\psi \rightarrow \phi \eta^\prime (\phi \rightarrow K^+K^-)$ &  $(1.97\pm 0.34)$  & 1.97 $\times 10^6$ \\
$J/\psi \rightarrow \omega \eta^\prime (\omega \rightarrow \pi^+\pi^-\pi^0) $ &  $(1.62 \pm 0.19)$  & 1.62$\times 10^6$  \\
$J/\psi \rightarrow \rho \eta^\prime (\rho \rightarrow \pi^+\pi^-) $ &  $(1.05 \pm 0.18)$  & $1.05\times 10^6$ \\
$J/\psi \rightarrow p\bar{p} \eta^\prime$ &  $(9.0\pm 4.0)$ &
9.0$\times 10^6$ \\ \hline
 Total &   & 60.74$\times 10^6$ \\\hline
\end{tabular}\label{tab:etap}}
\end{table}

\section{Hadronic decays of $\eta$ and $\eta^\prime$}

  Hadronic decays of $\eta$, $\eta^\prime \rightarrow \pi^+\pi^-
  \pi^0$ and $\pi^0\pi^0\pi^0$  can be utilized to extract $m_d -
  m_u$. The 3$\pi$-decay of $\eta$ or $\eta^\prime$ violates
  iso-spin invariance. In case of the $\pi^0\pi^0\pi^0$ system the
  two pions can have $I_{2\pi} = 0$, 1, 2 but coupling with the
  remaining pion to $I_{3\pi} =0 $ is only possible if $I_{2\pi}
  =1$. However, the $(\pi^0\pi^0)_{I=1}$ dose not exist due to requirement of Bose-Einstein statistics and as a
  consequence the decay $\eta(\eta^\prime) \rightarrow
  \pi^0\pi^0\pi^0$ has to violate isospin. In the case of the
  $\eta (\eta^\prime) \rightarrow \pi^+\pi^- \pi^0$ decay one can
  write~\cite{thesis}
\begin{equation}
(3\pi)_{I =0} = \sqrt{\frac{1}{3}} [ (\pi^+\pi^0)_{I=1}|\pi^-
\rangle - (\pi^+\pi^-)_{I=1}|\pi^0 \rangle +
(\pi^-\pi^0)_{I=1}|\pi^+ \rangle],
 \label{3pions}
\end{equation}
where $(\pi^+\pi^0)_{I=1}= \sqrt{\frac{1}{2}}
[|\pi^+\rangle|\pi^0\rangle - |\pi^0\rangle | \pi^+ \rangle]$,
$(\pi^+\pi^-)_{I=1}= \sqrt{\frac{1}{2}}
[|\pi^+\rangle|\pi^-\rangle - |\pi^-\rangle | \pi^+ \rangle]$, and
$(\pi^-\pi^0)_{I=1}= \sqrt{\frac{1}{2}}
[-|\pi^-\rangle|\pi^0\rangle + |\pi^0\rangle | \pi^- \rangle]$,
one can obtain the full wave function for the $3 \pi$ system as :
\begin{eqnarray}
(3\pi)_{I =0} &=& \sqrt{\frac{1}{6}} [  |\pi^+\rangle |\pi^0
\rangle |\pi^- \rangle - |\pi^0 \rangle |\pi^+ \rangle |\pi^-
\rangle - |\pi^+ \rangle |\pi^- \rangle |\pi^0 \rangle \nonumber \\
& & + |\pi^-\rangle |\pi^+ \rangle |\pi^0 \rangle - |\pi^- \rangle
|\pi^0 \rangle |\pi^+ \rangle + |\pi^0 \rangle |\pi^- \rangle
|\pi^+ \rangle ],
 \label{3wave}
\end{eqnarray}
which is always antisymmetric against any exchange of pions. In
particular, we have $C (3\pi)_{I =0} = - (3 \pi)_{I=0}$. While it
is $ C= +1$ for $\eta (\eta^\prime)$. Therefore the decay $\eta
(\eta^\prime) \rightarrow \pi^+\pi^- \pi^0$ violates $C$ or $I$.
On the other hand there exist a $G$ operator which is constructed
from the $C$ parity and isospin $I_2$ operators as $G = C e^{i\pi
I_2}$, and the decay $\eta (\eta^\prime) \rightarrow \pi^+\pi^-
\pi^0$ should also violate the $G$ parity.

  According to Sutherland's theorem,
electromagnetic contributions to the process are very
small~\cite{sutherland} and the decay is induced dominantly by the
strong interaction via the $u$, $d$ mass difference. Based on the
following two assumptions: (1) the decay $\eta^\prime \rightarrow
\pi^0\pi^+\pi^-$ proceeds entirely via $\eta^\prime \rightarrow
\eta \pi^+\pi^-$ followed by $\pi^0 - \eta$ mixing; (2) both decay
amplitudes are "{\it essentially constant}" over phase space on
the Dalitz plot, Gross, Treiman and Wilczek claimed
that~\cite{gross2}
\begin{equation}
r = \frac{\Gamma(\eta^\prime \rightarrow \pi^0 \pi^+\pi^-)}{
\Gamma(\eta^\prime \rightarrow \eta \pi^+\pi^-)} =
(16.8)\frac{3}{16} \left(\frac{m_d -m_u}{m_s}\right)^2,
 \label{rationeta}
\end{equation}
However recently Borasoy {\it et al.} claimed that the light quark
masses cannot be extracted from the
ratio~\cite{borasoy1,ref3pions} since the results from the full
chiral unitary approach are in plain disagreement with these two
assumptions. It turns out that more precise experimental data on
$\eta$ and $\eta^\prime$ decays are needed. An improvement of the
experimental situation is foreseen in the near future due to the
upcoming data from WASA-at-COSY~\cite{wasa}, MAMI-C~\cite{mami}
and KLOE~\cite{kloe}. Here, we would address the clean data
samples from $\psi$ decays at BES-III~\cite{bepcii}.
\begin{table}[htbp]
\centering
  \caption{The production of $\eta (\eta^\prime) \rightarrow 3 \pi$ decay at
BES-III in charmonium decays by assuming $10\times 10^9$ J$/\psi$
and $3.0\times 10^9$ $\psi(2S)$ events can be collected per year.}
{\begin{tabular}{@{}lll} \toprule Decay mode &
Branching fraction  & \#Events \\
& (\%) &  per year  \\
\hline
$\eta \rightarrow \pi^0 \pi^+\pi^- $ &  $(22.73\pm 0.28)$     & 14.3 $\times 10^6$ \\
$\eta \rightarrow \pi^0\pi^0\pi^0$ &  $(32.56\pm 0.23)$  & 20.5 $\times 10^6$ \\
$\eta^\prime \rightarrow \eta \pi^+\pi^- $ &  $(44.6 \pm 1.4 )$  & 27.2$\times 10^6$  \\
$\eta^\prime \rightarrow \eta \pi^0\pi^0  $ &  $(20.7 \pm 1.2)$  & $12.6\times 10^6$ \\
$\eta^\prime \rightarrow \pi^0 \pi^+\pi^-$ &  $(0.37^{+0.11}_{-0.09} \pm 0.04)$~\cite{cleoc} & 0.23$\times 10^6$ \\
$\eta^\prime \rightarrow \pi^0 \pi^0\pi^0$ &  $(0.154 \pm 0.026)$
& $0.09\times 10^6$ \\\hline
\end{tabular}\label{tab:etaphadronic}}
\end{table}

The $\eta/\eta^\prime \rightarrow 3\pi$ is an ideal laboratory for
testing ChPT. From a fit to the Dalitz plot density distribution
one can make precise determinations of the parameters that
characterize the decay amplitude or. One can choose two of the
pion energies ($T_+$,$T_-$,$T_0$) in the $\eta$ rest frame, two of
the three combinations of the two-pion masses squared
($m^2_{+-}$,$m^2_{-0}$,$m^2_{0+}$) also called ($s$,$t$,$u$). The
Dalitz plot distribution for the charged decay channel is
described by the following two variables:
\begin{eqnarray}
X &=& \sqrt{3}\frac{T_+ -T_-}{Q_c} =
\frac{\sqrt{3}(u-t)}{2m_{\eta}Q_c}, \nonumber \\
Y& =& \frac{3T_0}{Q_c}-1 = \frac{3[(m_{\eta}-m_{\pi^0})^2
-s]}{2m_{\eta}Q_c} - 1,
 \label{xyp}
\end{eqnarray}
where $Q_c = T_0 + T_+ + T_- = m_{\eta} - 2m_{\pi} -m_{\pi^0}$.
For the neutral decay channel it is convenient to use one fully
symmetrized coordinate:
\begin{equation}
Z = \frac{2}{3} \sum^{3}_{i=1} \left( \frac{3T_i}{Q_n}-1\right)^2
= X^2 +Y^2,
 \label{zp}
\end{equation}
with $Q_n = m_{\eta} - 3m_{\pi^0}$, in order to reflect  symmetry
in all Mandelstam variables.

 The squared absolute values of the two decay amplitudes are
 expanded around the center of the corresponding Dalitz plot
 for $\eta /\eta^\prime \rightarrow \pi^0 \pi^+\pi^-$ in order to obtain the Dalitz slope parameters~\cite{borasoyepja26} :
\begin{equation}
|A_c (X,Y)|^2 = |{\cal N}_c|^2 [ 1+ a Y + b Y^2 +c X+dX^2 +eXY+
...],
 \label{camplitudes}
\end{equation}
while for the decays into three identical particles Bose symmetry
dictates the form
\begin{equation}
|A_n (X,Y)|^2 = |{\cal N}_n|^2 [1+2\alpha Z + ...],
 \label{nampli}
\end{equation}
For the charged channel odd terms in $X$ are forbidden due to
charge conjugation symmetry. The parameters ($a$,$b$,$c$,$d$,$e$
and $\alpha$) can be obtained from fits to the observed Dalitz
plot density, and can be computed by the theory.

In table~\ref{tab:etaphadronic}, we estimate the produced number
of events for various hadronic $\eta$ and $\eta^\prime$ decay at
BES-III with one year's data taking. There are about 14 million
$\eta \rightarrow \pi^0\pi^+\pi^-$ and 0.23 million
$\eta^\prime\rightarrow \pi^0\pi^+\pi^-$ events produced each
year, respectively. By considering the detector coverage and
reconstruction efficiencies for charged and neutral tracks at
BES-III, the selection efficiency for $\eta/\eta^\prime
\rightarrow \pi^0\pi^+\pi^-$ mode is estimated to be
30\%~\cite{bepcii}, and is constant over the Dalitz plots. Thus
the expected observed number of events in the Dalitz plot is about
$N_{exp} = 4.2$ million for $\eta \rightarrow \pi^0\pi^+\pi^-$
decay or $N_{exp} = 0.07 $ million for $\eta^\prime \rightarrow
\pi^0 \pi^+\pi^-$ decay. At KLOE, about 6.6 million of $\eta
\rightarrow \pi^0\pi^+\pi^-$ decays should be selected after
considering the selection efficiency in 2.5 fb$^{-1}$
data~\cite{kloe}. We expect that the Dalitz parameter can be
measured almost at the same level of sensitivity for the $\eta$
decay at BES-III.

Some specific integrated asymmetries as defined in
reference~\cite{prllay72} are very sensitive in assessing the
possible contributions to $C$-violation in amplitudes with fixed
$\Delta I$. In particular left-right asymmetry tests $C$-violation
with no specific $\Delta I$ constraint. One can calculate the
asymmetry as
\begin{equation}
A_{LR} = \frac{N_+ -N_-}{N_++N_-},
 \label{lrasymm}
\end{equation}
where $N_+$ is the number of events for which $\pi^+$ has greater
energy than $\pi^-$, and $N_-$ is the number of events for which
the $\pi^-$ has the greater energy in the $\eta$ rest system. One
can also measure the quadrant asymmetry $A_Q$ and sextant
asymmetry $A_S$ as defined in reference~\cite{prllay72}. The
quadrant asymmetry is sensitive to an $I=2$ final state, while the
sextant asymmetry is sensitive to an $I =0$
$C$-invariance-violating final state~\cite{tdlee,nauen}. With 4.2
million $\eta \rightarrow \pi^0\pi^+\pi^-$ events at BES-III, the
sensitivity of these asymmetry could be $7.0\times 10^{-4}$, which
is one order lower than the current PDG
measurements~\cite{pdg2008}

For the neutral channel $\eta/\eta^\prime \rightarrow
\pi^0\pi^0\pi^0$ decay, assuming that the averaged selection
efficiency is 20\%, we should expect that about 4.1 million of
$\eta \rightarrow 3\pi^0$ and 0.02 million $\eta^\prime
\rightarrow 3 \pi^0$ decays can be selected in the Dalitz plots at
BES-III. With these high statistics and low background samples,
one can make more precise measurement of the slope parameter
$\alpha$ as shown in formula~\ref{nampli}. The best measurement of
$\alpha$ is $\alpha = -0.032 \pm 0.003$ which is from the Crystal
Ball at MAMI-C~\cite{mami2008} by using about 3 million $\eta$
decay events. The result is within errors compatible with the
result from KLEO, $\alpha = -0.027 \pm 0.004
(stat)^{+0.004}_{-0.006}(syst)$, based on 650 thousand
events~\cite{refalpha}. We expect more precision measurement will
be obtained at BES-III with 4.1 million of $\eta \rightarrow
3\pi^0$ events.

For the $\eta^\prime \rightarrow \eta \pi^+ \pi^-$ and
$\eta^\prime \rightarrow \eta \pi^0\pi^0$ decays, about 27 million
and 12 million decay events could be detected at BES-III each
year, respectively. Since the mass of $\eta^\prime$ meson is large
enough, in this case, the contributions of $\sigma$, $a_0$ and
$f_0(980)$ resonances and their interference on the Dalitz plots
dominate~\cite{beisert}. It will be very interesting to study the
$\eta \pi$ and $\pi\pi$ scattering in these decay modes at BES-III
with the huge data set.

\section{Rare decays of $\eta$ and $\eta^\prime$}


Decays of $\eta$ and $\eta^\prime$ mesons into a lepton-antilepton
pairs, $\eta/\eta^\prime \rightarrow l^+l^-$, represent a
potentially important channel to look for effects of new
physics~\cite{zphysc1982}. The dominant mechanism within the SM is
a second order electromagnetic process, additionally suppressed by
helicity conservation, involving two virtual photons $\eta
\rightarrow \gamma^* \gamma^*$, which is sensitive to the form
factor $F_{\eta \gamma^*\gamma^*}$ of the transition $\eta
\rightarrow \gamma^*\gamma^*$ with off-shell
photons~\cite{drell1959}. The imaginary part of the decay
amplitude can be uniquely related to the decays width of the $\eta
\rightarrow \gamma \gamma$ decay. The experimental value of the
$\Gamma (\eta \rightarrow \gamma \gamma )$ leads to a lower limit
(the unitarity bound) of the branching ratio: $BR(\eta \rightarrow
e^+e^-) \geq 1.7 \times 10^{-9}$ and $BR(\eta \rightarrow
\mu^+\mu^-) \geq 4.4 \times 10^{-6}$ when the real part of the
decay amplitude is neglected~\cite{zphysc1982,landrp1985}. The
measured branching fraction, $BR(\eta \rightarrow \mu^+\mu^-) =
(5.7 \pm 0.8)\times 10^{-6}$~\cite{pdg2008}, is consistent with
this limit.

  The real part of the amplitude of the $\eta \rightarrow e^+e^-$
  decay can be estimated using the measured value of $BR(\eta
  \rightarrow \mu^+ \mu^-)$~\cite{tom1994,pich1998,wise1992,masso1993}.
  The assumption that the ratio between Im and Re parts of the
  amplitudes for the $\eta \rightarrow e^+e^-$ and $\eta \rightarrow \mu^+\mu^-$
  is the same leads to the prediction $BR(\eta \rightarrow e^+e^-) \simeq 6\times 10^{-9}$.
The limit for $\eta \rightarrow e^+ e^-$ is much lower than for
other decays of $\pi^0$ and $\eta$ into lepton-antilepton pairs.
This makes the $\eta \rightarrow e^+e^-$ decay rate sensitive to a
possible exotic contribution. The best experimental upper limit
for the $BR(\eta \rightarrow e^+ e^- )$ comes from the CLEO-II as
listed in table~\ref{tab:rareeta}. By the way, the decays $\pi^0
\rightarrow e^+e^-$, $\eta \rightarrow \mu^+\mu^-$ and $e^+e^-$
are also important in order to estimate long range contribution to
the decay $K_L \rightarrow \mu^+ \mu^-$~\cite{pich1998}.

Recently, the KTeV experiment at Fermilab has made the first
observation of the decay $\pi^0 \rightarrow e^+ e^-$ to
be~\cite{ktev2006}:
\begin{equation}
BR(\pi^0 \rightarrow e^+e^-) = (7.49\pm 0.29\pm 0.25) \times
10^{-8},
 \label{pi0ll}
\end{equation}
which is 3 standard deviations higher than the theoretical
prediction~\cite{alex2007,alex20081,alex20082}.  It is interesting
to search for the other neutral pseudoscalar meson decays into
lepton pairs and to compare with theoretical predictions. These
probes will offer a way to study long-distance dynamics in the
Standard Model.
 At BESIII, leptonic decays $\eta/\eta^\prime \rightarrow e^+e^-$,
 $\mu^+\mu^-$, $e^+e^- e^+e^-$, $\mu^+\mu^- \mu^+\mu^-$ and
 $e^+e^- \mu^+\mu^-$ can be measured with sensitivities at
 $10^{-7}$ as listed in table~\ref{tab:rareeta} and ~\ref{tab:rareetap}.
 Most of the limits will be improved by one or two orders by using data with one
 year luminosity in J/$\psi$ and $\psi(2S)$ decays.
 We will also test C-invariance by improving the upper limits on
 the C-forbidden decays $\eta /\eta^\prime \rightarrow \pi^0
 e^+e^-$, $\pi^0 \mu^+\mu^-$, $3\gamma$ and $\eta^\prime \rightarrow \eta e^+e^-, \eta \mu^+\mu^-$.
 We will test CP-invariance by searching for $\eta /\eta^\prime
 \rightarrow \pi\pi$ and $4 \pi^0$ decays~\cite{mamicp}. The first
 experimental run at $J/\psi$ and $\psi(2S)$ peaks has been
 conducted at BES-III/BEPC-II. We expect to obtain more data so
 that the sensitivities listed in table~\ref{tab:rareeta} and ~\ref{tab:rareetap}.
can be reached.
\begin{table}[htbp]
\centering
  \caption{The sensitivity of $\eta$ rare and forbidden decays at BES-III.
The expected sensitivities are estimated by considering the
detector efficiencies for different decay mode at BES-III. We
assume no background dilution and the observed number of signal
events is zero. The BES-III limit refers to a 90\% confidence
level.  } {\begin{tabular}{@{}lll} \toprule Decay mode &
 Best upper limits  & BES-III limit\\
 & 90\% CL  & with one year's luminosity   \\
\hline
$\eta \rightarrow e^+e^- $ &  $7.7\times 10^{-5}$     & 0.7 $\times 10^{-7}$ \\
$\eta \rightarrow \mu^+\mu^-$ &  $(5.8\pm 0.8)\times 10^{-6} $  & 0.8 $\times 10^{-7}$ \\
$\eta \rightarrow e^+ e^- e^+ e^- $ &  $6.9 \times 10^{-5}$  & 0.9$\times 10^{-7}$  \\
$\eta \rightarrow \mu^+ \mu^- \mu^+ \mu^- $ &  $-$  & 1.5$\times 10^{-7}$  \\
$\eta \rightarrow \pi^+\pi^- e^+ e^- $ &  $(4.2 \pm 1.2)\times 10^{-4}$  & 1.3$\times 10^{-7}$  \\
$\eta \rightarrow \pi^+ \pi^- \mu^+ \mu^- $ &  $-$  & 1.4$\times 10^{-7}$  \\
$\eta \rightarrow \pi^0 \mu^+ \mu^- $ &  $5\times 10^{-6}$  & 1.5$\times 10^{-7}$  \\
$\eta \rightarrow \pi^0 e^+ e^- $ &  $4 \times 10^{-5}$  & 1.3$\times 10^{-7}$  \\
$\eta \rightarrow \pi^0\gamma $ &  $9\times 10^{-5}$  & 1.2$\times 10^{-7}$  \\
$\eta \rightarrow \pi^0\pi^0$ &  $3.5\times 10^{-4}$  & 1.8$\times 10^{-7}$  \\
$\eta \rightarrow \pi^+\pi^-  $ &  $1.3\times 10^{-5}$  & 0.8$\times 10^{-7}$  \\
$\eta \rightarrow \mu^+ e^- + \mu^- e^+ $ &  $6 \times 10^{-6}$  & 0.8$\times 10^{-7}$  \\
$\eta \rightarrow $ invisible  &  $6\times 10^{-4}$  & 60$\times 10^{-7}$  \\
\hline
\end{tabular}\label{tab:rareeta}}
\end{table}

\begin{table}[htbp]
\centering
  \caption{The sensitivity of $\eta^\prime$ rare and forbidden decays at
BES-III. The expected sensitivities are estimated by considering
the detector efficiencies for different decay mode at BES-III. We
assume no background dilution and the observed number of signal
events is zero. The BES-III limit refers to a 90\% confidence
level. } {\begin{tabular}{@{}lll} \toprule Decay mode &
Best upper limits & BES-III limit\\
&90\% CL  & with one year's luminosity   \\
\hline
$\eta^\prime \rightarrow e^+e^- $ &  $ 2.1\times 10^{-7} $     & 0.7 $\times 10^{-7}$ \\
$\eta^\prime \rightarrow \mu^+\mu^-$ &  $-$  & 0.8 $\times 10^{-7}$ \\
$\eta^\prime \rightarrow e^+ e^- e^+ e^- $ &  $-$  & 0.9$\times 10^{-7}$  \\
$\eta^\prime \rightarrow \mu^+ \mu^- \mu^+ \mu^- $ &  $-$  & 1.6$\times 10^{-7}$  \\
$\eta^\prime \rightarrow \pi^+\pi^- e^+ e^-$~\cite{cleoc} &  $(25^{+12}_{-9}\pm 5)\times 10^{-4}$  & 1.4$\times 10^{-7}$  \\
$\eta^\prime \rightarrow \pi^+ \pi^- \mu^+ \mu^- $~\cite{cleoc} &  $2.4\times 10^{-4}$  & 1.5$\times 10^{-7}$  \\
$\eta^\prime \rightarrow \pi^0 \mu^+ \mu^- $ &  $6.0 \times 10^{-5}$  & 1.6$\times 10^{-7}$  \\
$\eta^\prime \rightarrow \pi^0 e^+ e^- $ &  $1.4 \times 10^{-3}$  & 1.3$\times 10^{-7}$  \\
$\eta^\prime \rightarrow \pi^0\gamma $ &  $-$  & 1.2$\times 10^{-7}$  \\
$\eta^\prime \rightarrow \pi^0\pi^0$ &  $9.0 \times 10^{-4}$  & 1.9$\times 10^{-7}$  \\
$\eta^\prime \rightarrow \pi^+\pi^-  $ &  $2.9 \times 10^{-3}$  & 0.8$\times 10^{-7}$  \\
$\eta^\prime \rightarrow \mu^+ e^- + \mu^- e^+ $ &  $4.7 \times 10^{-4}$  & 0.8$\times 10^{-7}$  \\
$\eta^\prime \rightarrow $ invisible  &  $9.5\times10^{-4}$~\cite{cleoc}  & 140$\times 10^{-7}$  \\
$\eta^\prime \rightarrow \eta e^+e^- $ &  $2.4 \times 10^{-3}$     & 2.4$\times 10^{-7}$ \\
$\eta^\prime \rightarrow \eta \mu^+\mu^- $ &  $1.5 \times 10^{-5} $     &  3.1$\times 10^{-7}$ \\
\hline
\end{tabular}\label{tab:rareetap}}
\end{table}

 The $\eta (\eta^\prime) \rightarrow \pi^+\pi^- e^+ e^-$ decay is
interesting as test of $CP$ violation which is motivated by
corresponding test in $K_L$ decays. A recent prediction and
observations of an asymmetry were made in the distribution of
angles between the $\pi^+\pi^-$ and $e^+ e^-$ production planes in
$K_L \rightarrow \pi^+\pi^- e^+ e^-$ decay~\cite{ktevcp}. These
observations have triggered theoretical speculations that a
similar observation in $\eta \rightarrow \pi^+\pi^- e^+ e^-$ decay
might reveal unexpected mechanisms of $CP$ violation in flavor
conserving processes~\cite{bigi,gaodn}. At BES-III, we can observe
a $CP$-violating asymmetry in the $CP$ and $T$-odd variable
sin$\phi$cos$\phi$,
\begin{equation}
A = \frac{N_{sin\phi cos\phi >0.0} - N_{sin\phi
cos\phi<0.0}}{N_{sin\phi cos\phi >0.0} + N_{sin\phi cos\phi<0.0}},
 \label{lrasymm}
\end{equation}
where $\phi$ is the angle between the $e^+e^-$ and $\pi^+\pi^-$
planes in the $\eta$ center of mass system. This asymmetry
implies, with the mild assumption of unitarity to avoid exotic
$CPT$ violation~\cite{bigi}, time reversal symmetry violation. The
quantity sin$\phi$cos$\phi$ is given by $(\hat{n}_{ee}\times
\hat{n}_{\pi\pi})\cdot \hat{z}(\hat{n}_{ee}\cdot
\hat{n}_{\pi\pi})$, where the $\hat{n}$'s are the unit normals and
$\hat{z}$ is the unit vector in the direction of the $\pi\pi$ in
the $\eta$ center of mass system. The measured branching fraction
is $BR(\eta \rightarrow \pi^+\pi^- e^+e^-) = (4.2 \pm 1.2)\times
10^{-4}$~\cite{pdg2008}. About 26 thousand events are expected to
be produced at BES-III. Assuming that the efficiency is about
30\%, we expect that the sensitivity to measure the $CP$ asymmetry
is about 1\%. A measurement done with a sensitivity better than
$10^{-2}$ for the asymmetry will provide a stringent constraint
for new physics proposed by D.~N.~Gao~\cite{gaodn}.

Invisible decays of  $\eta$ and $\eta^\prime$ mesons  offer a
window into what may lie beyond the Standard
Model~\cite{Fayet:1979qi,Fayet:2006sp}. The reason is that apart
from neutrinos, the Standard Model includes no other invisible
final particles that these states can decay into.  It is such a
window that we intend to further explore by presenting here the
first experimental limits on invisible decays of the $\eta$ and
$\eta'$, which complement the limit of $2.7 \times 10^{-7}$
recently established in \cite{Artamonov:2005cu} for the invisible
decays of the $\,\pi^\circ$.

 Theories beyond the Standard Model generally include new physics, such as,
possibly, light dark matter (LDM) particles~\cite{Boehm:2003hm}.
These can have the right relic abundance to constitute the
nonbaryonic dark matter of the Universe, if they are coupled to
the SM through a new light gauge boson $U$~\cite{Fayet:1980ad}, or
exchanges of heavy fermions.  It is also possible to consider a
light neutralino with coupling to the Standard Model mediated by a
light scalar singlet in the next-to-minimal supersymmetric
standard model~\cite{Ellis:1988er}.

The BES-II Collaboration searched for the invisible decay modes of
$\eta$ and $\eta^\prime$ for the first time in \phett \mbox{}
using the 58 million $J/\psi$ events at BES II~\cite{invisible}.
They obtained limits on the ratio,
$\displaystyle\frac{\mathcal{B}(\eta(\eta^\prime) \to
  \text{invisible})}{\mathcal{B}(\eta(\eta^\prime) \to\gamma\gamma)}$.
The upper limits at the 90\% confidence level are $1.65\times
10^{-3}$ and $6.69\times 10^{-2}$ for
$\displaystyle\frac{\mathcal{B}(\eta \to
  \text{invisible})}{\mathcal{B}(\eta \to\gamma\gamma)}$ and
$\displaystyle\frac{\mathcal{B}(\eta^\prime \to
  \text{invisible})}{\mathcal{B}(\eta^\prime \to\gamma\gamma)}$,
respectively, corresponding to upper limits on the rates for $\eta
\rightarrow \text{invisible}$ of 6.0 $\times 10^{-4}$ and for
$\eta^\prime \rightarrow  \text{invisible}$ of 1.4$\times
10^{-3}$. CLEO-c has almost two times better limit than that from
BES-II for $\eta^\prime \rightarrow \text{invisible}$ decay as
shown in table~\ref{tab:rareetap}~\cite{cleoc}. At BES-III, the
sensitivity of the invisible decays will be improved by two orders
of magnitude with high statistics data as listed in
table~\ref{tab:rareeta} and ~\ref{tab:rareetap}.

\section{Summary}

A mini-review of $\eta$ and $\eta^\prime$ physics at BES-III has
been done. With one year's luminosity data, about 63 million
$\eta$ decays and 61 million $\eta^\prime$ decays can be collected
at BES-III.  The hadronic decays of $\eta$ and $\eta^\prime$ can
be studied, especially parameters for the Dalitz decays of $\eta$
could be extracted from about 4.2 million $\eta$ events, so that
the prediction of ChPT will be tested.  The rare and forbidden
decays of $\eta$ and $\eta^\prime$ could be reached at BES-III,
and the sensitivity of these decay is at the level of $10^{-7}$.
New physics beyond the standard Model can be probed at low energy.
At BES-III, $\eta$ and $\eta^\prime$ decay samples are clean, and
the event topology is simple so that events can be easily
reconstructed. More fruit results will be seen from BES-III
experiment.

\section*{Acknowledgments}
The author would like to thank D.~N.~Gao and M.~Z.~Yang for useful
discussions. This work is supported in part by the National
Natural Science Foundation of China under contracts Nos.
10521003,10821063, the 100 Talents program of CAS, and the
Knowledge Innovation Project of CAS under contract Nos. U-612 and
U-530 (IHEP).




\end{document}